# Improving Child Speech Recognition and Reading Mistake Detection by Using Prompts

*Lingyun Gao, Cristian Tejedor-Garcia, Catia Cucchiarini, Helmer Strik*

Centre for Language Studies, Radboud University, the Netherlands

{lingyun.gao, cristian.tejedorgarcia, catia.cucchiarini, helmer.strik}@ru.nl

## Abstract

Automatic reading aloud evaluation can provide valuable support to teachers by enabling more efficient scoring of reading exercises. However, research on reading evaluation systems and applications remains limited. We present a novel multimodal approach that leverages audio and knowledge from text resources. In particular, we explored the potential of using Whisper and instruction-tuned large language models (LLMs) with prompts to improve transcriptions for child speech recognition, as well as their effectiveness in downstream reading mistake detection. Our results demonstrate the effectiveness of prompting Whisper and prompting LLM, compared to the baseline Whisper model without prompting. The best performing system achieved state-of-the-art recognition performance in Dutch child read speech, with a word error rate (WER) of 5.1%, improving the baseline WER of 9.4%. Furthermore, it significantly improved reading mistake detection, increasing the F1 score from 0.39 to 0.73.

**Index Terms**: child speech recognition, reading mistake detection, prompts, Whisper, large language models, LLM, prompting

## 1. Introduction

Reading aloud is regularly assessed in primary education as a measure of reading skills, but this form of assessment is labor-intensive [1] and error-prone, as teachers must manually evaluate read speech, identify mistakes, and assign scores. Automating this process could improve efficiency, which is important as declining literacy rates highlight the need for innovative solutions [2]. Advances in large pretrained ASR systems have made speech-based assessment more feasible [3], yet most ASR models are designed for adults [4, 5] and struggle with child speech recognition [6, 7, 8, 9].

Downstream reading evaluation and especially reading mistake detection presents even greater challenges [7, 10, 11, 12], as most ASR techniques tend to prioritize transcribing correct speech, thereby neglecting reading mistakes. It has been observed that most publicly available large pretrained ASR models struggle with incorrectly spoken speech. ASR models with a language model as a decoder, such as Whisper [5], excel at transcribing correctly spoken words, but frequently fail to detect reading mistakes. In contrast, models with a Connectionist Temporal Classification (CTC) decoder like wav2vec2-CTC [13] are better at identifying reading mistakes, but struggle with overall transcription accuracy.

Concerning the evaluation of read-aloud speech, there is another important observation. Evaluation of read-aloud speech is inherently multimodal, as teachers assess speech by referencing familiar texts and using their understanding of common reading mistakes [14, 15]. However, few studies integrate multimodal input (e.g., combining audio and text) or prior knowledge into this task. Two main approaches enhance ASR transcription using multimodal input. The first involves ASR models that integrate text prompts. For instance, the Whisper model uses text prompts as contextual cues for speech transcription. Kany & Trouvain [16] highlight its effectiveness in transcribing fillers in child read-aloud speech, while Harmsen et al. [8] examine its application to reading word lists. Although Whisper often corrects reading mistakes rather than directly transcribing them, its confidence scores aid in detecting misread words. These findings indicate that modifying prompts can adjust Whisper's output style to support downstream tasks. However, Whisper's ability to incorporate prior knowledge remains limited [17]. Instruction-tuned large language models (LLMs) may present a more effective alternative. The second approach utilizes LLMs with prompts to refine ASR transcriptions. Southwell et al. [18] enhance child conversation recognition through LLM-based hypothesis rescoring, while Tur et al. [19] use LLM prompting to generate refined ASR hypotheses. However, the potential of using Whisper and LLMs with prompts for recognizing child read speech, and especially detecting reading mistakes, remains largely unexplored.

In this work, we present a novel contribution by investigating the use of prompts with Whisper and large language models to integrate diverse prior textual information for child speech recognition and reading mistake detection. Given the scarcity of resources in this area, our study particularly emphasizes methodological innovations. The research questions (RQs) we address are: RQ1. To what extent can Whisper with prompts and LLM with prompts enhance performance in child speech recognition?

RQ2. To what extent can these two methods improve reading mistake detection?

## 2. Method

### 2.1. Data

This study uses read speech data from the Jasmin-CGN Corpus [20], featuring recordings of 71 Dutch primary school pupils (ages 6–13, 35 female, 36 male) reading three stories aloud at their mastery levels. The dataset includes orthographic and phonemic annotations, as well as reading mistake annotations for the first story [21]. The data normalization process includes removing punctuation and capitalization, as well as standardizing Dutch words with varying formats.

The test dataset consists of 2.05 hours of speech of children reading the first story, encompassing 14,251 reading attempts. These include 11,322 correctly read words, 615 incorrect words or word fragments, and attempts with inaudible words. Additionally, a validation dataset is provided, which includes 55-second audio recordings of five sentences with reading mistakes selected from the third reading story of a single speaker, paired with orthographic transcriptions.

## 2.2. Evaluation Metrics

In this work, we analyze substitution, deletion, and insertion errors made by children during reading aloud. To detect these mistakes, we use force alignment with SCTK[1], following the loose criteria of [22], where a detected mistake is considered valid if its predicted label and position match the ground truth. We assess child speech recognition performance using Word Error Rate (WER), which quantifies transcription accuracy based on these errors, with lower WER indicating better performance. For reading mistake detection, we evaluate Precision (correctly identified mistakes among predictions), Recall (detected mistakes among actual mistakes), and F1-score (harmonic mean of Precision and Recall).

## 2.3. Prompting Methods

We address our RQs by constructing two prompting pipelines (Figure 1): prompting Whisper and prompting LLM. In the prompting Whisper pipeline (Figure 1.a), we examine whether Whisper with prompts can improve speech transcription with reading mistakes and enhance downstream reading mistake detection. This is achieved by prompting Whisper models with text containing generated mistakes while exploring optimal settings for this approach. We use Whisper models as our speech recognition module as they support text prompt transfer styles of transcriptions[2] and appear to be the state-of-the-art (SOTA) models for Dutch child speech recognition [7]. Then, we explore the potential of integrating knowledge from read text and other CTC model transcriptions to Whisper transcriptions by prompting LLM and its impact in reading mistake detection, as shown in Figure 1.b.

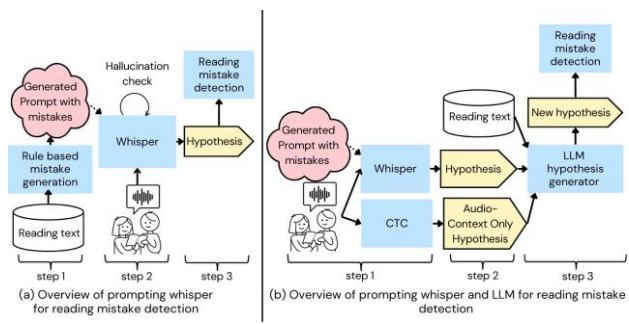

Figure 1: *Overviews of prompting Whisper pipeline (a) on the left side and prompting LLM pipeline (b) on the right side.*

### 2.3.1. Prompting Whisper Pipeline

We form a three-step reading mistake detection pipeline for prompting Whisper as shown in Figure 1.a. In Step 1, we apply reading mistake categories [23, 24] and use a rule-based mistake generation algorithm to insert random mistakes before the target word in the reading text (e.g., "het is zondag" becomes "h het it is zon zondag"). This mistake-inserted text is then used as the prompt for Whisper. We test three configurations: (1) reading text with 10% mistakes (one mistake per ten words), (2) read text with 100% mistakes (3) four irrelevant read text sentences with 300% mistakes. Since Whisper has a limited token capacity for prompts, reading text containing a story with 100% mistakes has nearly reached the maximum token size. These settings help assess the impact of mistake proportion and resource of additional context on Whisper hypothesis transferring. Next, in Step 2, the Whisper model processes the audio together with the generated prompt to transcribe the input speech. We perform hallucination check and then use it for reading mistake detection by comparing the number of tokens in the hypothesis with the original reading text. On average, child audio contains approximately 8% reading mistakes [21]. Based on this observation, we establish a heuristic hallucination threshold, which is confirmed through the validation dataset. In our method, hypotheses that are 20% longer or 5% shorter than the read text are flagged as hallucinations and replaced with a standard Whisper transcription (without prompt). Finally, in Step 3, we align the new hypotheses generated with different Whisper models with the orthographic transcription and the read text to explore the effectiveness of our methods with different Whisper models on improving child speech recognition and the best model on downstream tasks.

### 2.3.2. Prompting LLM Pipeline

We apply a multi-step reading mistake detection pipeline involving Whisper and the LLM, as shown in Figure 1.b. In Step 1, we obtain the Whisper hypothesis using the optimal prompting Whisper method (irre. 300% mistake) and we generate hypothesis transcriptions from a speech foundation model with CTC decoder, which serves as the audio-context-only hypothesis. This paper employs wav2vec2, further pretrained and fine-tuned on Dutch adult speech [25], due to its superior performance in Dutch child speech recognition and insertion mistake detection compared to other CTC models[7]. In Step 2, text prompts for the LLM are formulated by integrating: (1) the read text for correcting name entities and infrequent words (2) examples in common reading mistakes in children [23] to determine plausible mistakes, (3) improve Whisper mistake hypothesis by checking CTC hypotheses and the read text (e.g.,improving insertion, keeping phonetically similar substitution mistakes or reverting to the reference word) and examples. The LLM is then instructed to generate an improved output based on modifying the Whisper hypothesis. Two input-output configurations are evaluated: (1) the LLM receives the reading text and both hypotheses as input and produces a refined hypothesis, and (2) the LLM is provided with forced alignments for (read text, Whisper hypothesis) and (read text, wav2vec2 hypothesis) pairs and generates a forced alignment for the (read text, new hypothesis) pair. In Step 3, we obtain the new hypothesis either directly from the LLM response or by extracting it from the output alignment. We then evaluate the effectiveness of the generated new hypothesis from different LLM models in improving child speech recognition and to determine the best-performing model for the downstream reading mistake detection task.

Regarding the selection of LLMs, we choose GPT-4o-mini, which achieves the best Dutch language generation and understanding tasks among relatively small (<100B) LLM models [26], as our baseline LLM. In addition to GPT-4o-mini[3], we evaluated Llama 3.3-70B-Instruct[4], and DeepSeek-R1-Distill-Llama 70B[5], as they are similarly sized open-source models. Llama 3.3-70B-Instruct has shown comparable performance in few-shot Dutch language tasks with GPT-4o-mini, while DeepSeek-R1-Distill-Llama 70B exhibits potential for superior reasoning capabilities compared to GPT-4o-mini.

---

[1] https://github.com/usnistgov/SCTK, accessed in January 2025
[2] https://cookbook.openai.com/examples/whisper_prompting_guide, accessed in January 2025
[3] https://platform.openai.com/docs/models, accessed in December 2024
[4] https://huggingface.co/meta-llama/Llama-3.3-70B-Instruct, accessed in February 2025
[5] https://huggingface.co/deepseek-ai/DeepSeek-R1-Distill-Llama-70B, accessed in Feb. 2025

## 3. Results

We experimented with a single A100 GPU, WhisperX [27], huggingface pipeline and OpenAI API, and DeepSeek API. The code and prompts used in this project can be found online[6].

### 3.1. Prompting Whisper

Table 1 shows the child speech recognition performance of the prompting Whisper method under various prompt configurations for Whisper Large v2 and Large v3. The prompting Whisper pipeline using irrelevant text with 300% mistakes can effectively improve child speech recognition for both Whisper models. For Whisper v2, compared to the "No Prompt" setting, providing the model with a prompt containing the read text decreases performance. However, providing the model with irrelevant text with 300% mistakes proves effective, achieving the lowest WER of 8.1%. Whisper Large v3 performs comparably to Whisper Large v2 in child read speech recognition without prompting, and each setting of the prompting Whisper method has a similar effect on the two models. Whisper Large v3 exhibits significantly more hallucinations, causing most transcriptions to revert to the baseline during hallucination checks, which slightly improves its performance in read-text prompting settings.

Table 1: *Comparison of WER using different prompt settings for Whisper Large v2 and Whisper Large v3*

| Whisper prompt settings | WER: v2 | WER: v3 |
|---|---|---|
| no prompt | 9.4 | 9.3 |
| read text with 10% mistakes | 13.1 | 9.7 |
| read text with 100% mistakes | 11.3 | 9.8 |
| irrelevant text with 300% mistakes | **8.1** | 8.9 |

### 3.2. Prompting LLM

Table 2 summarizes WER results for different prompt settings using GPT-4o-mini in the pipeline, with or without the prompting Whisper method. With optimal prompting from Whisper, WER is 8.1%. When the LLM is prompted with the Whisper hypothesis to generate a new hypothesis directly ("+LLM: hypothesis"), performance declines. However, providing the LLM with alignment hypotheses—constructed from Whisper transcriptions and the read text—and asking it to refine them ("+LLM: alignment hypothesis") significantly improves child speech recognition, reducing WER to 5.1%. The same trend is observed without prompting Whisper, but its optimal setting remains less effective than incorporating prompting Whisper.

Table 2: *Comparison of WER using different prompt settings for LLM with and without prompting Whisper*

| prompting LLM settings | LLM + Whisper | LLM + prompting Whisper |
|---|---|---|
| No prompt | 9.4 | 8.1 |
| +LLM:hypothesis | 9.8 | 9.1 |
| +LLM:alignment hypothesis | 7.9 | **5.1** |

Table 3 compares WER scores across three LLMs under two prompt settings, using the optimal prompting Whisper method versus the no-prompt setting. In the "LLM: hypothesis" setting, WER worsens for all models, with Llama 70B showing the least degradation at 9.0%. Conversely, the "LLM: alignment hypothesis" setting significantly improves WER across all models, with Llama 70B achieving the best performance at 5.1%. This suggests that the proposed prompting method is generalizable across LLMs of similar size. Furthermore, the relative improvement of GPT-4o-mini and LLaMA 70B correlates with their Dutch understanding task scores[26], indicating that WER improvements align with each model's Dutch proficiency.

Table 3: *Comparison of WER across three LLM models*

| prompting LLM settings | GPT-4o-mini | LLama 70b | DeepSeek-R1-distill 70b |
|---|---|---|---|
| no LLM | 8.1 | 8.1 | 8.1 |
| LLM:hypothesis | 9.1 | 9.0 | 9.4 |
| LLM:alignment hypothesis | **5.1** | 7.6 | 7.7 |

### 3.3. Reading Mistake Detection

Table 4 evaluates the effectiveness of different settings for prompting Whisper and prompting LLM in detecting reading mistakes, with Whisper Large v2 and GPT-4o-mini. Baseline Whisper (without prompting) achieves an overall F1=0.39, excelling in deletions (F1=0.71), but struggling with insertions (F1=0.27) and substitutions (F1=0.48). Adding read text with 10% mistakes lowers overall performance (F1=0.36), while using read text with 100% mistakes significantly improves it (F1=0.58), especially for insertions (F1=0.68) and substitutions (F1=0.63), though deletions drop sharply. The highest performance (F1=0.61) is achieved with irrelevant text containing 300% mistakes, improving insertions (F1=0.67), maintaining deletions (F1=0.70), but slightly reducing substitutions (F1=0.45).

Table 4: *Comparison of reading mistake Detection Performance on precision, recall and F1 for different prompting Whisper settings and prompting LLM settings based on whisper with irrelevant text + 300% mistakes*

| Prompt settings | All mistakes | | | Ins | Sub | Del |
|---|---|---|---|---|---|---|
| | Precision | Recall | F1 | F1 | F1 | F1 |
| Whisper without prompt | 0.54 | 0.31 | 0.39 | 0.27 | 0.48 | 0.71 |
| +read text 10% mistakes | 0.37 | 0.35 | 0.36 | 0.29 | 0.42 | 0.35 |
| +read text 100% mistakes | 0.50 | **0.71** | 0.58 | **0.68** | **0.63** | 0.36 |
| +irre. text 300% mistakes | **0.58** | 0.65 | **0.61** | 0.67 | 0.45 | 0.70 |
| Whisper + irrelevant text as prompt | | | | | | |
| +LLM:hypothesis | 0.72 | 0.32 | 0.45 | 0.41 | 0.38 | **0.73** |
| +LLM:hypothesis alignment | **0.76** | **0.70** | **0.73** | **0.76** | **0.75** | 0.71 |

These results show that both the "read text with 100% mistakes" and "irrelevant text with 300% mistakes" settings effectively improve reading mistake detection, particularly for insertion mistakes. This suggests that prompting Whisper with a high-error prompt (>100% errors) effectively enhances the Whisper model's ability to capture reading mistakes. However, the "read text with 100% mistakes" setting substantially improves substitution detection while reducing precision and significantly impairing deletion detection. This suggests that this setting improves ASR transcriptions for incorrectly read words, but introduces excessive errors, thereby impairing the detection of deletion mistakes.

With the optimal prompting-Whisper method, incorporating LLMs can further boost performance with the hypothesis alignment settings. The best overall performance is observed (Table 4) with the LLM using hypothesis alignment, which achieves an F1 score of 0.73 for all mistakes. This setting demonstrates balanced improvements across all metric and mistake types, with insertion, substitution, and deletion F1 scores of 0.76, 0.75, and 0.71, respectively. These findings suggest that combining Whisper with irrelevant prompts for transcription and LLMs with hypothesis alignment for refinement is highly effective in improving reading mistake detection. However, using an LLM with hypothesis as input/output lowers the F1 score, achieving higher

---
[6]https://github.com/Lingygao/reading-mistake-detection-with-prompt, accessed in May 2025

precision, but lower recall. This highlights the importance of alignment for the effectiveness of prompting the LLM method.

### 3.4. Frequent ASR Transcription Error Analysis

This section examines frequent ASR substitution and deletion errors across our proposed methods, exploring how these errors change with different approaches to reveal patterns of improvement and challenges. Table 5 categorizes the most frequent ASR substitution errors (>6 occurrences) into seven types: breaking long words, misrecognizing named entities, misrecognizing multisyllabic words as phonetically similar words (with similar vowels or similar consonants), misrecognizing monosyllabic words, inserting or deleting often silent/unvoiced letters (e.g., e/n/t/h), and other substitutions, to provide a insightful view of substitution error patterns in the optimal prompting Whisper and prompting LLM methods. Compared to the baseline Whisper method, the optimal prompting Whisper method (using irrelevant text as a prompt) introduces new substitution errors, such as breaking long words (e.g., 'cameraploeg' → 'camera ploeg'), similar-vowel errors ('anneleen' → 'annelien'), and increased insertion/deletion of e/n/t/h, particularly n ('toonden' → 'toonde'), which shows a general decrease in accuracy. Other Whisper prompting variations, which use reading text with generated mistakes, show similar error patterns, but sometimes misrecognize words as artificially generated mistakes in the prompt (e.g., 'lekke' misrecognized as 'lekkke'), suggesting that prompting with reading text with generated mistakes may misguide Whisper by treating the generated mistakes as a reference.

Table 5: *Categorization of the most frequent ASR substitution errors (>6 occurrences) across Baseline, prompting Whisper with irrelevant text, and prompting LLM with hypothesis alignment*

|                      | Baseline (%)  | Prompting Whisper (%) | Prompting LLM (%) |
|----------------------|---------------|-----------------------|-------------------|
| Break long words     | 0 (0.0%)      | 8 (4.65%)             | 0 (0.0%)          |
| Name entity error    | 7 (7.53%)     | 7 (4.07%)             | 0 (0.0%)          |
| Similar consonant    | 16 (17.20%)   | 15 (8.72%)            | 0 (0.0%)          |
| Similar vowel        | 5 (5.38%)     | 18 (10.47%)           | 0 (0.0%)          |
| Mono-syllable        | 23 (24.73%)   | 26 (15.12%)           | 12 (17.91%)       |
| Delete/insert e/n/t/h| 7 (7.53%)     | 47 (27.33%)           | 10 (14.93%)       |
| Other substitutions  | 35 (37.63%)   | 51 (29.65%)           | 45 (67.16%)       |
| **Total**            | **93 (100%)** | **172 (100%)**        | **67 (100%)**     |

In contrast, the optimal Prompting LLM method (using hypothesis alignment for both input and output) effectively reduces substitution errors in top frequent long words and named entities, which may largely rely on knowledge from the reading text. It also reduces errors in phonetically similar words, monosyllabic words, and substitutions caused by insertions or deletions of e/n/t/h, indicating that LLMs utilize Dutch phoneme-grapheme knowledge. These findings suggest that LLMs integrate reading text and linguistic knowledge to improve transcription accuracy. However, when prompted with raw ASR hypotheses and asked to revise them, LLMs show only slight improvements in substitution errors and continue to struggle with long-text alignment. Similarly, using hypothesis alignment as input while generating a new hypothesis produces results similar to direct hypothesis prompting, suggesting LLMs may favor correcting pre-existing errors over aligning mistakes accurately.

The five most frequent ASR deletion errors in the Whisper baseline, ['de' (47), 'en' (30), 'ze' (30), 'dat' (25), 'een' (25)], align with common self-repetition and short-word mistakes reported in [21]. The Prompting Whisper setting, with over 100% generated mistakes, improves recognition and reduces deletions. In irrelevant text settings, deletions shift to ['de' (26), 'en' (16), 'een' (12), 'ggg:noise' (12), 's' (12)], though the 10% mistake setting worsens deletion errors. Compared to optimal prompting Whisper, the prompting LLM method slightly reduces monosyllable word deletions. These findings suggest that prompting Whisper effectively transfers a transcription style that captures self-repetition and short-word insertion mistakes more accurately than the baseline Whisper, while further prompting with LLM improves recognition, possibly by integrating the captured mistakes from the CTC model.

## 4. Discussion and Conclusion

In this work, we combined Whisper with prompts and LLMs with prompts to enhance child speech recognition and reading mistake detection, particularly in settings where resources are limited. By addressing two RQs, we evaluated the effectiveness of these methods in improving child speech recognition and reading mistake detection.

To address RQ1, namely to what extent Whisper with prompts and LLM with prompts can enhance performance in child speech recognition, we analyzed word recognition performance as measured by WER of the two prompting methods in various settings and models. The prompting Whisper method using irrelevant text with generated 300% mistakes effectively enhanced child speech recognition for both Whisper models. The best result was achieved with Whisper Large v2, reducing WER to 8.1% from the baseline 9.4% (Table 1). The prompting LLM method with hypothesis alignment further improved recognition, with GPT-4o-mini achieving the lowest WER of 5.1% (Table 2).

Regarding RQ2, namely to what extent these prompting methods can improve reading mistake detection, we applied Whisper Large v2 and GPT-4o-mini and analyzed metrics including precision, recall and F1 scores for reading mistake detection of the two prompting methods across different settings. As seen in the rows at the top of Table 4, both the "read text with 100% mistakes" and "irrelevant text with 300% mistakes" settings improved Recall and F1, but the irrelevant text setting improved three metrics. The highest F1 score at 0.61 was achieved by the "irrelevant text setting", up from the baseline 0.39. As seen in the bottom part of Table 4, the prompting LLM method with hypothesis alignment setting further enhanced detection, reaching the highest F1-score at 0.73.

Besides, the ASR error analysis showed that prompting Whisper with high-error prompts (more than 100% mistakes in prompt text) helped detect more reading mistakes, particularly self-repetitions and short-word insertions. However, this also led to increased hallucinations and reduced word accuracy. Interestingly, using irrelevant text as a prompt, instead of the actual read text, proved more effective. This likely prevented the model from using the prompt as a spelling reference. The prompting LLM method further improved performance by leveraging infrequent words, Dutch phoneme knowledge, and CTC transcriptions which provided additional mistake predictions. Our results suggested broad applicability across LLMs, but the method required extra forced-alignment tools. In our proposed method, the LLM must be prompted with aligned text and must produce aligned output, making the process computationally intensive. Also, its ability to handle long speech segments and additional input text is constrained by the LLM's token limit.

In future work, we will address the sensitivity of the prompting LLM method, which depends on external tools and LLM variability, affecting reliability. Furthermore, biases towards human mistakes in speech and LLM models may impact child speech recognition across tasks, particularly in low-resource settings. Therefore, future research will focus on enhancing robustness.

## 5. Acknowledgments

This publication is part of the project Responsible AI for Voice Diagnostics (RAIVD) with file number NGF.1607.22.013 of the research programme NGF AiNed Fellowship Grants which is financed by the Dutch Research Council (NWO).